\pdfoutput=1 %

\documentclass[aps,prl,twocolumn,superscriptaddress,groupedaddress]{revtex4}
\usepackage[T1]{fontenc} %
\usepackage{amsmath,amssymb,amsfonts,amsthm}
\usepackage{graphicx}
\usepackage{enumitem}
\usepackage{verbatim}
\usepackage[percent]{overpic}
\usepackage{rotating}
\usepackage{hyperref}
\usepackage{array}
\usepackage{mathtools}
\usepackage{lmodern}
\usepackage[normalem]{ulem}
\usepackage{braket}
\usepackage{tensor}
\usepackage{dsfont}
\usepackage{bm}
\usepackage{tikz}
\usepackage{fancyhdr}
\usepackage{mathrsfs}
\usetikzlibrary{decorations.pathmorphing}
\usepackage{CJKutf8}
\usepackage{simplewick}

\usepackage{ascmac}

\usetikzlibrary{positioning}
\usetikzlibrary{calc,through,backgrounds}
\usepackage[font=small,labelsep=none]{caption}
\usepackage{subfig}

\theoremstyle{definition}
\newtheorem{definition}{Definition}[section]
\newtheorem{theorem}{Theorem}[section]

\newcommand{\kako}[1]{\left( #1 \right)}
\newcommand{\kagikako}[1]{\left[ #1 \right]}
\newcommand{\ts}[1]{ _{\text{#1}} }
\newcommand{\Bigkako}[1]{\Big( #1 \Big)}
\newcommand{\biggkako}[1]{\bigg( #1 \bigg)}
\newcommand{\Bigkagikako}[1]{\Big[ #1 \Big]}

\newcommand{\erfi}{\text{erfi}}

\allowdisplaybreaks

\DeclareMathOperator{\Tr}{Tr}

\newcommand{\dd}{\text{d}}

\newcommand{\id}{\mathds{1}}
\newcommand{\sx}{\mathsf{x}}

\newcommand{\ii}{\mathsf{i}}

\usepackage[scr=boondoxo]{mathalfa}	%
\newcommand{\qew}{\mathscr{q}}				%

\newcommand{\lb}{\left}						%
\newcommand{\rb}{\right}					%
\newcommand{\nn}{\nonumber}					%
\newcommand{\bd}{\boldsymbol}				%
\newcommand{\mc}{\mathcal}					%
\newcommand{\tp}{\otimes}					%
\newcommand{\iu}{\mathrm{i}}		%
\newcommand{\ec}{\mathrm{e}}	%

\begin{document} 

\title{Instant Extraction of Non-Perturbative Tripartite Entanglement}
\author{Diana M\'endez Avalos}
\email[]{g7_menav18@ens.cnyn.unam.mx}
\affiliation{
Centro de Nanociencias y Nanotecnología
Univiersidad Nacional Autónoma de México (CNyN-UNAM)
Km 107 Carretera Tijuana-Ensenada, C.P. 22800, Ensenada, B.C.\ Mexico}
\author{Kensuke Gallock-Yoshimura}
\email{kgallock@uwaterloo.ca}
\affiliation{Department of Physics and Astronomy, University of Waterloo, Waterloo, Ontario, N2L 3G1, Canada}
\author{Laura J. Henderson}
\email[]{l7henderson@uwaterloo.ca}
\affiliation{Department of Physics and Astronomy, University of Waterloo, Waterloo, Ontario, N2L 3G1, Canada}
\affiliation{Institute for Quantum Computing, University of Waterloo, Ontario, N2L 3G1, Canada}
\author{Robert B. Mann}
\email[]{rbmann@uwaterloo.ca}
\affiliation{Department of Physics and Astronomy, University of Waterloo, Waterloo, Ontario, N2L 3G1, Canada}
\affiliation{Institute for Quantum Computing, University of Waterloo, Ontario, N2L 3G1, Canada}

\begin{abstract}
We consider the problem of
extracting tripartite entanglement
through single local instantaneous interactions of a  separable target system A-B-C with a scalar field. 
We find, non-perturbatively, that tripartite entanglement is easily extracted in this scenario, in strong contrast to   bipartite extraction, which is not possible due to a no-go theorem.  The tripartite entanglement is of the GHZ-type, and an optimal value of the coupling exists that admits maximal extraction.
\end{abstract}

\maketitle

\clearpage
\noindent\emph{Introduction}$\quad$  
The entanglement of quantum fields has been an area of significant attention over the past few decades, with areas of interest in quantum information\cite{Peres:QuantumInfo,Lamata:Entanglement} and metrology \cite{Ralph:connectivity}, quantum energy teleportation \cite{doi:10.1143/JPSJ.78.034001,Hotta:2011xj}, the AdS/CFT correspondence \cite{Ryu:AdSCFT}, black hole entropy \cite{Solodukhin2011,Brustein2005} and the black hole information paradox \cite{Preskill:1992tc,Mathur:2009hf,almheiri2013black,Braunstein:2009my,Mann:2015luq,Louko2014firewall,Luo:2017}.

The entanglement present in the vacuum state of a quantum field was originally shown by Valentini \cite{Valentini1991nonlocalcorr} and later by Reznik \textit{et.\ al.\ }\cite{reznik2003entanglement} to be able to be swapped to a pair first-quantised particle detectors, even if the two detectors are not in causal contact throughout the duration of the interaction.  The extracted entanglement was subsequently   shown  to be distillable into Bell pairs \cite{reznik2005violating}.  Since then, there has been a significant amount of research\cite{Steeg2009,olson2012extraction,pozas2015harvesting,smith2016topology,Vacuum_Harvesting_experiment2,kukita2017harvesting,Simidzija.Nonperturbative,Simidzija2018no-go,henderson2018harvesting,ng2018AdS,henderson2019entangling,cong2019entanglement,Tjoa2020vaidya,cong2020horizon,Xu:2020pbj} 
investigating the process of entangling a pair of two-level particle detectors, known as Unruh-DeWitt (UDW) detectors\cite{Unruh1979evaporation,DeWitt1979}, through localised, time dependent interactions with the vacuum state of quantum scalar field.  This process has become to be known as \textit{entanglement harvesting} \cite{salton2015acceleration}.

The process of swapping field entanglement to particle detectors is not limited to bipartite entanglement, but can be extended to multipartite entanglement as well, of which much less is known.
 Silman and Reznik demonstrated that a finite-duration interaction of $N$ UDW detectors with the vacuum of a scalar field yields a reduced density matrix containing  $N$-partite entanglement between the detectors that can in principle be distilled to that of a $W$-state\cite{SilamReznik:Wstate}.  Some time later Lorek \textit{et.\ al.\ } used Gaussian quantum mechanics to show that three harmonic oscillators in a $(1+1)$-dimensional cavity can extract genuine tripartite entanglement following  interaction with the scalar field vacuum state  even if the detectors remained spacelike separated\cite{Lorek2014tripartite}. 
 Furthermore,  they found that in this context it was easier to harvest tripartite entanglement than bipartite entanglement.  More recently, it was shown using perturbation theory that three initially uncorrelated UDW detectors in $(3+1)$-dimensional Minkowski space are able to harvest tripartite entanglement from the scalar field.  Outside of the context of entanglement harvesting, it was found that acceleration leads to a degradation in the tripartite entanglement of initially entangled detectors \cite{Hwang:pra2011,Szypulski:2021}.

A nonperturbative calculation of the final state of the three detectors following the interaction with field can be carried out by assuming the   detectors couple instantaneously with the field, which is modelled using Dirac-delta switching.  If each detector couples only once to the field with delta-switching, bipartite entanglement harvesting has been shown to be  impossible\cite{Simidzija.Nonperturbative}: at least one detector will need to couple twice\cite{Simidzija2018no-go} or the the dectectors must switch in a
temporal superposition \cite{Henderson2020temporal}.  
More recently it has been shown that a third party, with a detector that also couples to the field with delta-switching, is able to prevent any bipartite correlations from forming between the two detectors \cite{Sahu:prd2022}, 
indicating the existence of multipartite entanglement between the three parties.  

Here we demonstrate that  harvesting tripartite entanglement with delta-switching is considerably easier than the bipartite case: we can nonperturbatively extract genuine tripartite entanglement  when each of the detectors switches only once. 
This is 
in striking contrast with the bipartite case, where at least one detector must  switch more than once to allow for any entanglement between the detector pair.  
An optimal value of the coupling exists for maximally 
harvesting the tripartite entanglement, which is of the GHZ-type, having no residual bipartite entanglement upon tracing over any one of the three detectors. \\

\noindent\emph{Harvesting Protocol}$\quad$
\label{sec:ThreeDet}
Consider three static UDW detectors $A, B$, and $C$ in an $(n+1)$-dimensional Minkowski spacetime. 
Each detector $D\in \{ A, B, C \}$ has a spatial shape specified by a smearing function $F_D(\bm{x}-\bm{x}_D) \in \mathbb{R}$, where $\bm{x}_D$ is the position of center of mass, and instantaneously locally interacts with a quantum scalar field $\hat \phi(\sx)$ via the interaction Hamiltonian
\begin{align}
    \hat H\ts{I}(t)
    &=  
    \sum_{D\in \{A,B,C\}}
    \lambda_D 
    \eta_{D} \delta (t-T_{D})
    \hat \mu_D(t) \nn\\
    &\hspace{1cm} \otimes 
    \int \dd^n x\,F_D(\bm{x}-\bm{x}_D)
    \hat \phi \big( \bd{x}(t) \big). %
\end{align}
that we refer to as  delta-switching \cite{Simidzija.Nonperturbative}.
Here $\lambda_D$ is the coupling constant, $\eta_{D} \delta (t-T_{D})$ is the delta switching with a strength $\eta_{D}$ and  $T_{D}$ is the time when detector-$D$ interacts with the field.
$\hat \mu_D(t)$ is the monopole moment of detector-$D$, which acts on different Hilbert spaces depending on $D$. 
For instance, $\hat \mu_A(t)=\hat m_A(t) \otimes \id_{B}\otimes \id_{C}$, and similarly for detectors $B$ and $C$, where 
\begin{align}
    \hat m_D(t)&=
    e^{ \ii \Omega_D t } \ket{1_D}\bra{0_D}
    +
    e^{ -\ii \Omega_D t } \ket{0_D}\bra{1_D}, 
\end{align}
with detector-$D$'s ground $\ket{0_D}$ and excited $\ket{1_D}$ states, and an energy gap $\Omega_D \in \mathbb{R}$ between them. 

Without loss of generality, we assume that the detectors turn on in the following order
\begin{align}
    T_A \leq T_B \leq T_C
    \label{eq:SwitchingOrder}
\end{align}
in which case the time-evolution operator in the interaction picture
$\hat U\ts{I}=
    \mathcal{T} \exp
    \kagikako{-\ii \int_{\mathbb{R}} \dd t\,
        \hat H\ts{I}(t)} $
(with $\mathcal{T}$ the time-ordering symbol) is nonperturbatively given by  
\begin{align}
    \hat U\ts{I}&=
    \ec^{
    \kagikako{
        \hat \mu_{C}(T_{C}) \otimes \hat Y_{C}(T_{C})
    } }
   \ec^{
    \kagikako{
        \hat \mu_{B}(T_{B}) \otimes \hat Y_{B}(T_{B})
    }} \ec^{ \kagikako{
        \hat \mu_{A}(T_{A}) \otimes \hat Y_{A}(T_{A})
    }}
     \label{eq:expanded unitary}
\end{align}
where 
\begin{align}
    \hat Y_D(t)
    &\coloneqq
    -
    \ii \lambda_D \eta_{D}
    \int \dd^n x\,F_D(\bm{x}-\bm{x}_D)
    \hat \phi \big( \bm{x}(t) \big) 
\end{align}
is the smeared field operator at  detector $D$'s location.

We next introduce two quantities 
\begin{align}
    \Theta_{D,E}&\coloneqq 
    -\ii \bra{0} [ \hat Y_{D}(T_D), \hat Y_{E}T_E) ] \Ket{0}  \label{eq:Comm}\\
    \omega_{D,E}&\coloneqq
    \dfrac12 \bra{0} \{ \hat Y_{D}(T_D), \hat Y_{E}(T_E) \} \ket{0} \label{eq:AntiComm}  %
\end{align}
which are respectively the 
vacuum expectation values of the commutator, $\Theta_{D,E}$, and anticommutator, $\omega_{D,E}$ for $D,E\in\{A,B,C\}$ with $D\ne E$. 

If two detectors  are causally disconnected then $\Theta_{D, E}=0$ and they cannot communicate.
The anticommutator $\omega_{D, E}$, 
on the other hand, is nonzero even when detectors cannot send and receive signals. 
In fact, the extracted entanglement using causally disconnected detectors directly comes from $\omega_{D, E}$ \cite{PhysRevD.104.125005}.

Suppose the field is massless and the initial state of the total system is 
\begin{align}
    \hat \rho_0&= \ket{0_{A} 0_{B} 0_{C} }\bra{0_{A} 0_{B} 0_{C}}\otimes \ket{0}\bra{0},
\end{align}
where $\ket{0}$ is the Minkowski vacuum state of the field. 
The final state $\hat \rho_{ABC}$ of the detectors can be obtained by tracing out the field degrees of freedom of the final state of the sytsem: $\Tr_\phi [ \hat U\ts{I} \hat \rho_0 \hat U\ts{I}^\dag ]$. 
The density matrix $\hat \rho_{ABC}$ in the basis $\{ \ket{0_{A}0_{B}0_{C}}, \ket{0_{A}0_{B}1_{C}}$, $\ket{0_{A}1_{B}0_{C}}, \ket{1_{A}0_{B}0_{C}}$, $\ket{0_{A}1_{B}1_{C}}$, $\ket{1_{A}0_{B}1_{C}}$, $\ket{1_{A}1_{B}0_{C}}, \ket{1_{A}1_{B}1_{C}} \}$ can be written as (see Supplementary Material for the derivation)
\begin{align}
    \hat \rho_{ABC}
    &=
        \left[
        \begin{array}{cccccccc}
        r_{11} &0 &0 &0 &r_{15} &r_{16} &r_{17} &0  \\
        0 &r_{22} &r_{23} &r_{24} &0 &0 &0 &r_{28}  \\
        0 &r_{23}^* &r_{33} &r_{34} &0 &0 &0 &r_{38}  \\
        0 &r_{24}^* &r_{34}^* &r_{44} &0 &0 &0 &r_{48}  \\
        r_{15}^* &0 &0 &0 &r_{55} &r_{56} &r_{57} &0  \\
        r_{16}^* &0 &0 &0 &r_{56}^* &r_{66} &r_{67} &0  \\
        r_{17}^* &0 &0 &0 &r_{57}^* &r_{67}^* &r_{77} &0  \\
        0 &r_{28}^* &r_{38}^* &r_{48}^* &0 &0 &0 &r_{88} 
        \end{array}
        \right] \label{eq:rhoABC}
\end{align}
where the matrix elements $r_{i,j}$ are functions of the quantities
\begin{subequations}
\begin{align}
    f_{D}&=
    \exp
    \kako{
        -\dfrac{1}{2} 
        \int \dd^n k\,|\beta_{D} (\bm{k})|^2
    }, \\
    \Theta_{D,E}&=
    \ii
    \int \dd^n k\,
    \Bigkako{
        \beta_{D}^*(\bm{k}) \beta_{E}(\bm{k})
        -
        \beta_{D}(\bm{k}) \beta_{E}^*(\bm{k})
    } , \\
    \omega_{D,E}&=
    -\dfrac{1}{2} 
    \int \dd^n k\,
    \Bigkako{
        \beta_{D}^*(\bm{k}) \beta_{E}(\bm{k})
        +
        \beta_{D}(\bm{k}) \beta_{E}^*(\bm{k})
    }, \\
    \beta_{D}(\bm{k})&\coloneqq
    -\dfrac{\ii \lambda_D \eta_{D}}{ 2 \sqrt{ 2 |\bm{k}| } }
    \tilde{ F }_D^* (\bm{k}) 
    \ec^{ \ii ( |\bm{k}| T_{D} - \bm{k}\cdot \bm{x}_{D} ) }, \label{eq:beta}
\end{align}
\label{eq:MatrixFunctions}
\end{subequations}
and where 
\begin{align}
    \tilde{F}_D(\bm{k})&=
    \int \dfrac{ \dd^n x }{ \sqrt{ (2\pi)^n } }
    F_D(\bm{x}) \ec^{\ii \bm{k}\cdot \bm{x}}
\end{align}
is the Fourier transform of the  smearing function, $F_D(\bm{x})$.
Once the shape of detectors $F_D(\bm{x})$ is specified, we can calculate each element in the density matrix.

With the density matrix of the final state of the three detectors fully calculated, we will now consider the resulting tripartite entanglement between the three detectors.  We choose the $\pi$-tangle as our measure of tripartite entanglement \cite{Ou:pitangle}, which puts a lower bound on the genuine tripartite entanglement in the mixed state $\hat{\rho}_{ABC}$.  It is defined as
\begin{equation}
    \pi\coloneqq\frac{1}{3}\lb(\pi_{A}+\pi_{B}+\pi_{C}\rb)
\end{equation}
with
\begin{equation}
    \pi_A\coloneqq\mathcal{N}_{A(BC)}^2-\mathcal{N}_{A(B)}^2-\mathcal{N}_{A(C)}^2 
\end{equation}
and similarly for subsystems $B$ and $C$.  The negativities of the subsystems are
\begin{align}
    \mathcal{N}_{A(BC)} &\coloneqq ||\hat \rho_{ABC}^{T_A}||-1  \label{eqn:NegativityTri}\\
    \mathcal{N}_{A(B)} &\coloneqq ||(\Tr_C[\hat \rho_{ABC}])^{T_A}||-1  \label{eqn:NegativityBi}
\end{align}
where $T_A$ is the partial transpose with respect to detector $A$ and $||\cdot||$ is the trace norm, with similar definitions for detectors $B$ and $C$.
We will refer to these as the tripartite and bipartite negativity respectively. In all scenarios we consider the
no-go theorem \cite{Simidzija.Nonperturbative} ensures
the latter quantity is zero, and so the $\pi$-tangle provides a measure of genuine tripartite entanglement.\\

\noindent\emph{Instant Tripartite entanglement harvesting} 
\label{sec:Results} $\quad$
We will now calculate the $\pi$-tangle of three identical detectors in   $(3+1)$-dimensional Minkowski spacetime 
 so that the  switching strength $\eta$,  coupling constant, $\lambda$, and
 energy gap $\Omega$ are the same for each. The latter assumption is not required since the elements of the density matrix 
 only depend on the gap via phases, which cancel out if the detectors switch only once.
 
We shall take the Gaussian smearing function 
\begin{equation}
    F_D(\bd{x})=\exp\lb(-\frac{(\bd{x}-\bd{x}_D)^2}{2\sigma^2}\rb),    
\end{equation}
with characteristic width $\sigma$
to describe the shape of each detector,
allowing for exact computation of $f_D$ from  \eqref{eq:MatrixFunctions}. 

\begin{figure}[h]
    \centering
    \includegraphics[width=\linewidth]{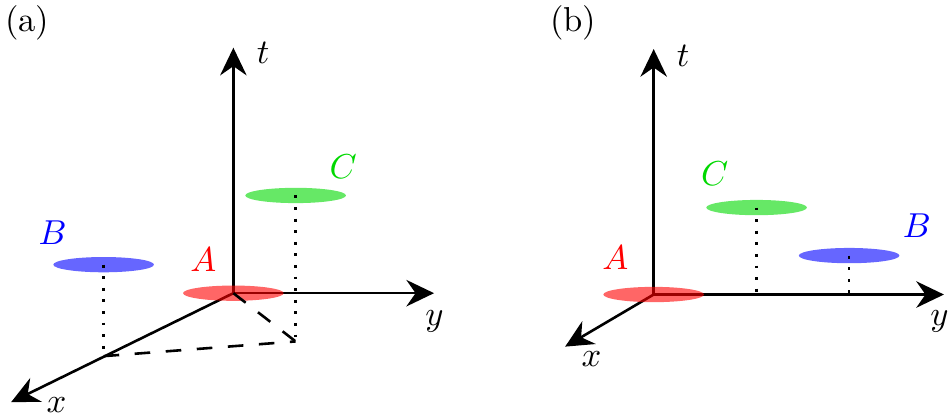}\\
    \caption{
    ~Spacetime diagram illustrating (a) the triangle configuration, and (b) the line configuration. }
    \label{fig:spacetime diagram}
\end{figure}

\begin{figure}[ht]
	\centering
	\includegraphics[width=\linewidth]{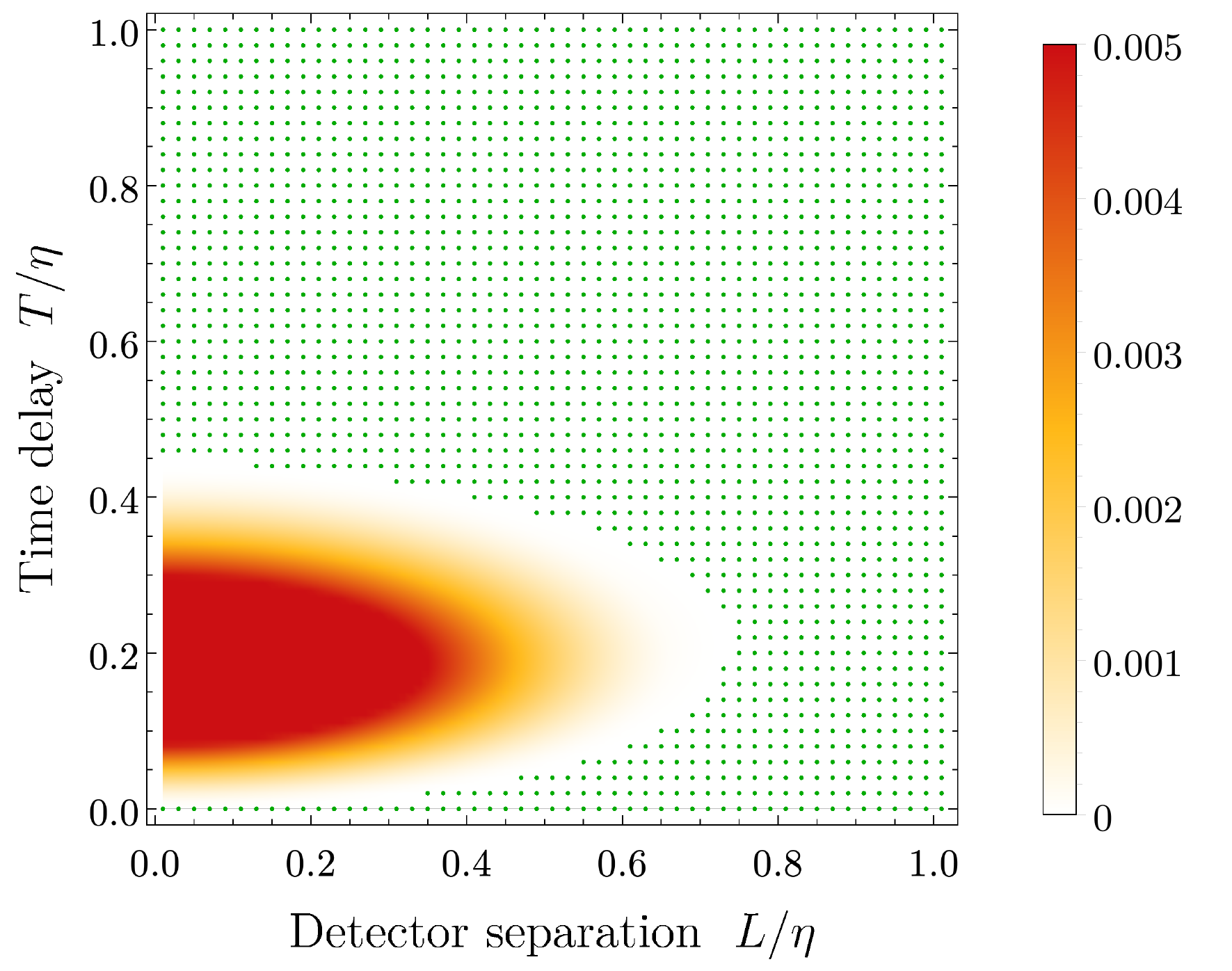}
	\caption{%
		~ The $\pi$-tangle as a function of the side length, $L$, and time delay, $T$, for three detectors arranged in an equilateral triangle that each couple to the field once with Dirac $\delta$ switching.  The detectors have an energy gap   $\Omega\eta=1$,  width   $\sigma=\eta$, and  coupling   $\lambda=10$.  The green dots indicate the regions of the parameter space where the $\pi$-tangle is zero.
	}
	\label{fig:piVsTandL}
\end{figure}

First, as shown in Fig. \ref{fig:spacetime diagram}(a), we will consider a configuration where the three detectors are placed  at the vertices of an equilateral triangle with side length $L$,  coupling to the field at times $T_A=0$, $T_B=T$ and $T_C=2T$.
We plot the $\pi$-tangle of this configuration as a function of $T$ and $L$
for $\lambda=10$
in Fig. \ref{fig:piVsTandL}. We see that the three detector system has genuine tripartite entanglement, after each detector couples to the field just once through delta-switching.  This is in contrast to bipartite entanglement between any two detectors, which is forbidden by the no-go theorem\cite{Simidzija.Nonperturbative}.

\begin{figure}[ht]
	\centering
	\includegraphics[width=\linewidth]{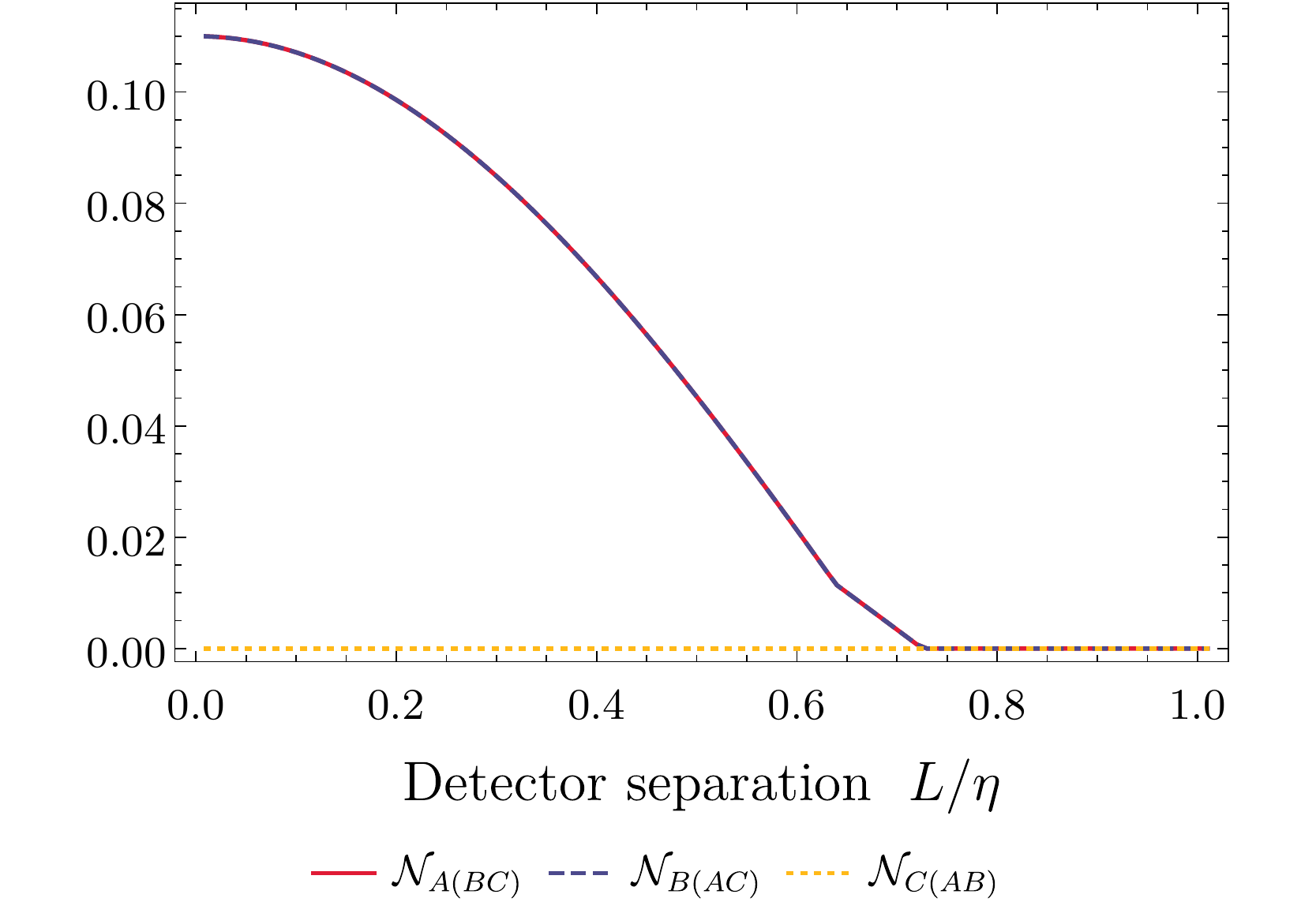}
	\caption{%
		~ The tripartite negativities as a function of the side length, $L$, for three detectors arranged in an equilateral triangle that each couple to the field once with Dirac delta-switching.  Two of the three tripartite negativities are non-zero (and equal), which results in a positive $\pi$-tangle.  
		All of the bipartite negativities are zero. The time between switches is $T=0.25\eta$, the detectors have an energy gap of $\Omega\eta=1$,  width   $\sigma=\eta$ and  coupling    $\lambda=10$.  
	}
	\label{fig:3Neg}
\end{figure}

The positive $\pi$-tangle is strictly a result of the positive tripartite negativities $\mc{N}_{A(BC)}$ and $\mc{N}_{B(AC)}$, as shown in Fig. \ref{fig:3Neg}.  In other words, detector $A$ is entangled with the $(BC)$ subsystem and detector $B$ is entangled with the $(AC)$ subsystem.  However, detector $C$ is not entangled with the $(AB)$ subsystem.  We also confirm that that the bipartite negativities between the detectors are zero (as required by the no-go theorem\cite{Simidzija.Nonperturbative}), meaning no pair of detectors are entangled. Consequently, the tripartite entanglement is of the GHZ-type  and not of the  W-type.

As expected, a third detector cannot be used to avoid the no-go theorem since the coupling of the third detector will act as a displacement operator on the field, moving it into a coherent state \cite{Simidzija.Nonperturbative}. 

The non-zero tripartite entanglement can be better understood by interpreting it from the perspective of subsystems.  The negativity $\mc{N}_{A(BC)}$ quantifies the entanglement between detector $A$   with the subsystem $(BC)$.  From this perspective, the first subsystem $A$ switches once, and then the second subsystem $(BC)$ switches twice, once at the location of detector $B$ and once at the location of detector $C$.  This type of switching is analogous to the BAA-type switching described in \cite{Simidzija2018no-go},
which is complex enough to allow the two systems to become entangled.  Similarly, the negativity $\mc{N}_{B(AC)}$ quantifies the entanglement between detector $B$ and the subsystem $(AC)$.  The switching order (Eq.\ \eqref{eq:SwitchingOrder}) means that the first subsystem $(AC)$ interacts with the field (at the location of detector $A$), then the second subsystem $B$ interacts with the field, and finally the first  subsystem interacts again with the field (at the location of detector $C$), which is analogous to the ABA-type switching that also allows for the systems to become entangled.  Following this logic, the negativity $\mc{N}_{C(AB)}$, quantifies the entanglement between the subsystem $(AB)$ and detector $C$.  The switching order of these systems is analogous to the AAB-type switching, which prohibits the systems from becoming entangled.

\begin{figure}[ht]
	\centering
	\includegraphics[width=\linewidth]{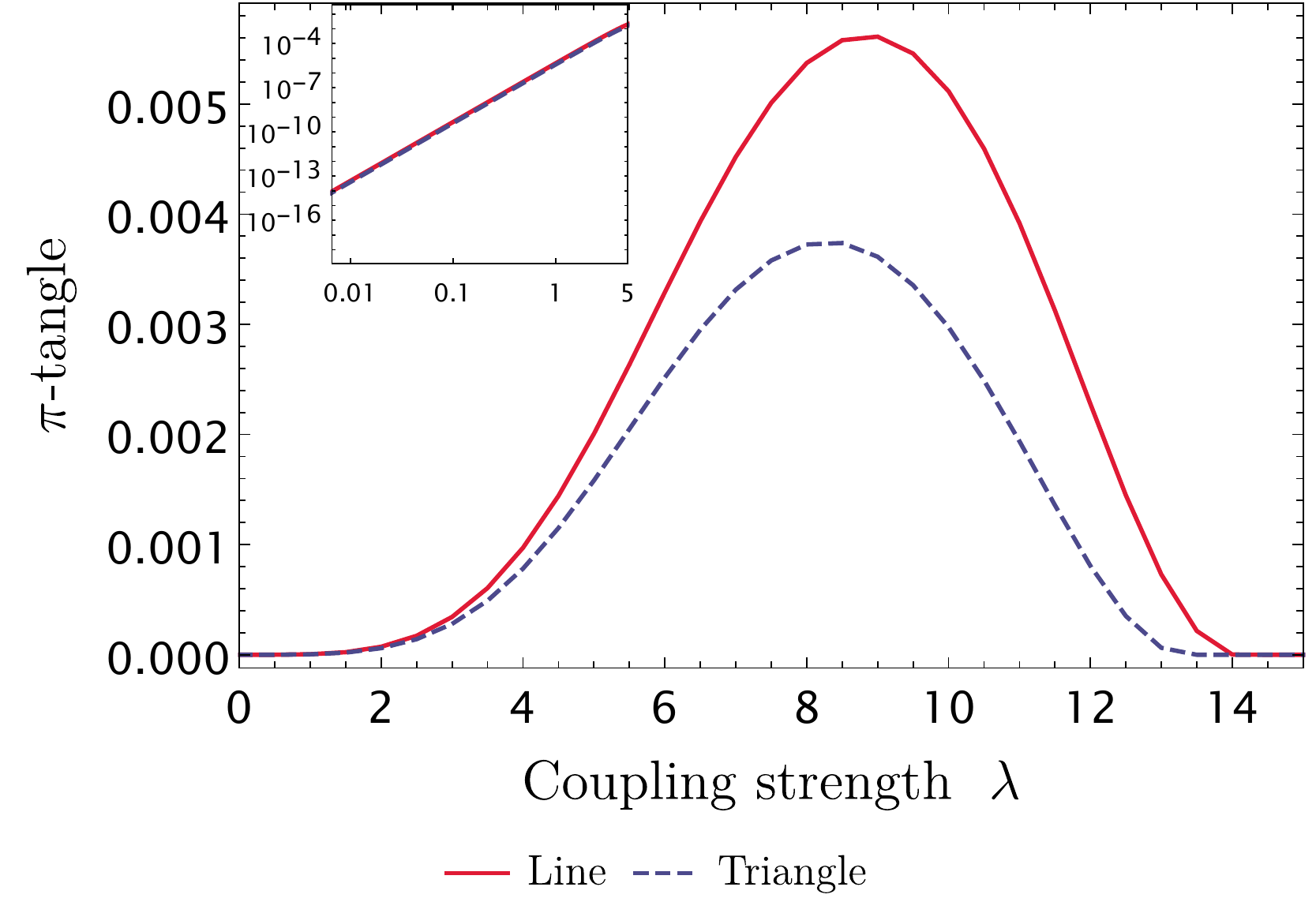}
	\caption{%
		~ A comparison of the $\pi$-tangle for three detectors arranged an equilateral triangle with side length $L=0.4\eta$ and in an line of length $L_{AB}=0.4\eta$ that each couple to the field once with Dirac delta-switching as a function of the coupling strength, $\lambda$.  The time between switches is $T=0.25\eta$ and the detectors have an energy gap of $\Omega\eta=1$ and a width of $\sigma=\eta$.
	}
	\label{fig:piVslambda}
\end{figure}

In Fig. \ref{fig:piVslambda}, we compare the equilateral triangle configuration [Fig. \ref{fig:spacetime diagram}(a)] with one 
where the detectors are arranged in a straight line with detector $C$ in the middle [Fig. \ref{fig:spacetime diagram}(b)], equidistant from detectors $A$ and $B$. Setting 
the distance between detectors $A$ and $B$ on the line to equal the side length of the equilateral triangle, we find that the $\pi$-tangle is significantly larger in the case of the line for all values of the coupling constant where tripartite entanglement harvesting is possible.  Furthermore, the $\pi$-tangle is non-zero for larger values of the coupling constant 
in the linear configuration.

Both of these effects are due to the reduced distance between detector $C$ and the other two detectors in the line configuration as opposed to the triangle.  The reduced distance means there is stronger signaling between detector $C$ and detectors $A$ and $B$, which leads to an increase in the resulting tripartite entanglement in the linear  configuration. We have checked this over a broad range of separations and time delays and found this general trend to hold. More explicitly, the final state of the three detector system depends on the quantities (Eq.\ \eqref{eq:MatrixFunctions}) which only depend on the relative spacetime positions of at most two detectors, and so the overall configuration of the detectors only matters insofar as it changes their relative pairwise distances.

We also find that the qualitative behaviour of the $\pi$-tangle as a function of the coupling strength is the same.  When the coupling strength is small $(\lambda\lesssim5)$, the $\pi$-tangle has quartic growth with increasing $\lambda$ as shown in the inset of Fig. \ref{fig:piVslambda}.  In this regime, the evolution of the detectors, and hence their final state, $\hat{\rho}_{ABC}$, can be described as a perturbative expansion in even powers of the coupling strength. Hence to lowest order, the matrix elements 
and $\pi$-tangle
will respectively be quadratic and quartic in $\lambda$.  As the coupling constant increases out of the pertubative regime, the $\pi$-tangle reaches a maximum and then falls off exponentially to zero. In other words, there is an optimal value of $\lambda$ for maximally harvesting tripartite entanglement. We have checked this for other configurations and find it to be a general feature.

\begin{figure}[ht]
	\centering
	\includegraphics[width=\linewidth]{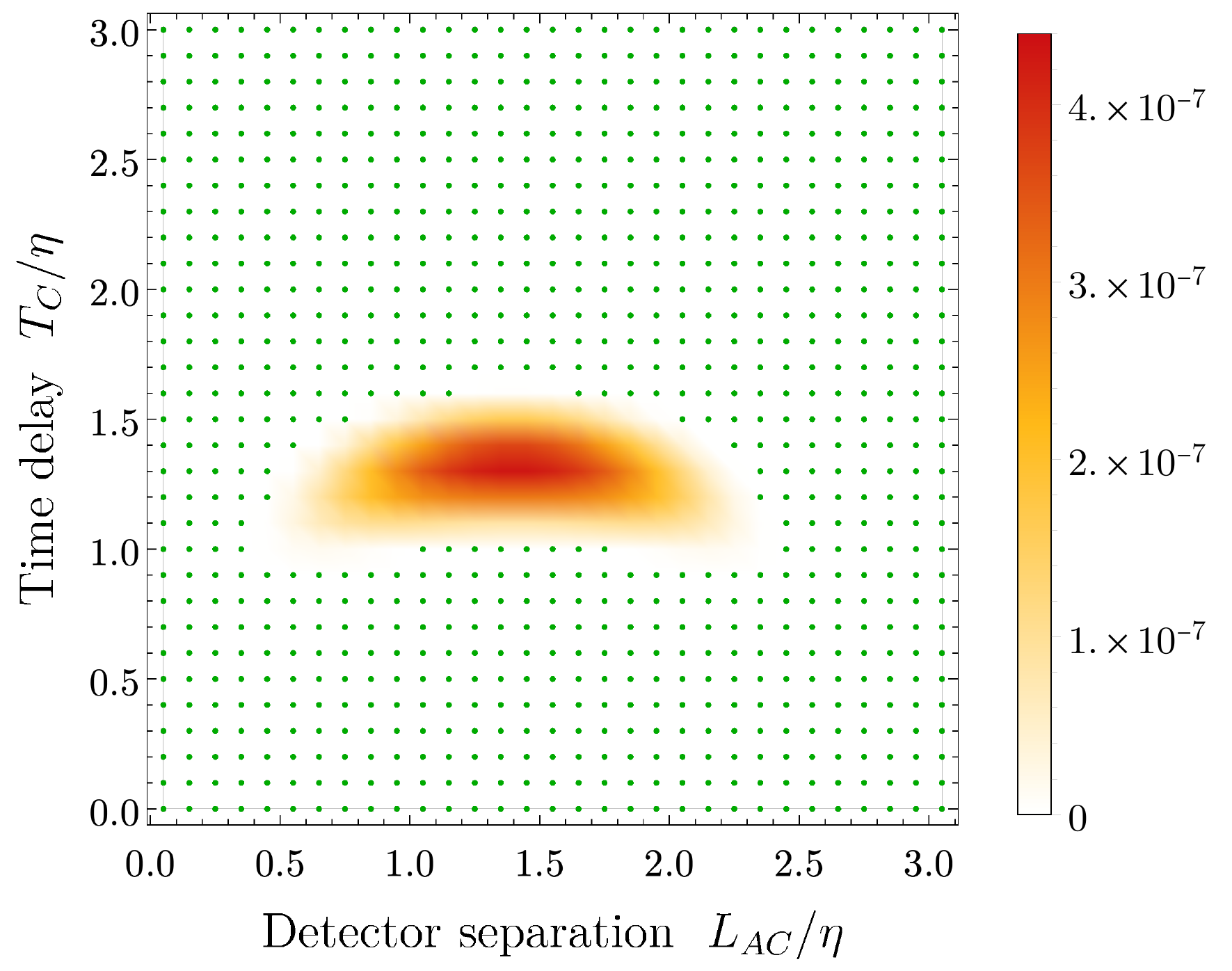}
	\caption{%
		~ The $\pi$-tangle for three detectors arranged in an line that each couple to the field once with Dirac $\delta$ switching as a function of switching time of detector $C$ and the distance of detector $C$ from $A$.  Detector $A$ and $B$ switch at $T_A=T_B=0$ and are separated by a distance of $L_{AB}=2.8\eta$. The detectors have an energy gap of $\Omega\eta=1$ and a width of $\sigma=\eta$ and the coupling constant is $\lambda=2.5$.  The green dots indicate the regions of the parameter space where the $\pi$-tangle is zero.
	}
	\label{fig:ABatSameTime}
\end{figure}

Finally, we explore the effects of signaling on the  tripartite entanglement.
We  first note that when two detectors switch at the same time, the commutator $\Theta_{A,B}=0$,   so they do not signal to each other.  We find in this case that three detectors can still have a positive $\pi$-tangle even if detectors $A$ and $B$ do not signal to each other; however the both must be able to signal to detector $C$. This is illustrated in Fig.  \ref{fig:ABatSameTime}, where  for the linear cnofiguration (Fig. \ref{fig:spacetime diagram}(b)), with 
$T_A=T_B=0$ and 
the distance between detectors $A$ and $B$ fixed to $L_{AB}=2.8\eta$,
we plot the $\pi$-tangle
as a function of $T_C\ge0$ and $L_{AC}$.
We observe that the $\pi$-tangle is positive only when detector $C$ is between $A$ and $B$ $(L_{AC}<L_{BC})$ and that it is maximum on the overlapping  lightcones of detectors A and B.   In other words, the $\pi$-tangle will be maximized if detector $C$ is positioned so that it maximizes its communication with both detectors $A$ and $B$.  Furthermore   the $\pi$-tangle is zero when detector $C$ is outside of the light code of only one of detectors $A$ or $B$,  we conclude that the $C$ must be able to communicate with both detectors in order for them to become tripartite entangled.

Reversing the situation so that the first detector $A$ switches followed by detectors $B$ and $C$ switching simultaneously, the $\pi$-tangle remains zero, which further illustrates that entanglement is only possible if the first two detectors are able to communicate with the third.

We emphasize that despite a significant amount of overlap in the smearing functions, the two detectors do not communicate since $\Theta_{AB}=0$. Indeed, when the smearing function of the detectors has a standard deviation of $\sigma=\eta$, if the distance between detectors $A$ and $B$ is increased beyond $L_{AB}=2.8\eta$, the $\pi$-tangle will be zero for the region of parameter space we explored.\\

\noindent\emph{Conclusion}
\label{sec:Con}
$\quad$
In strong contrast to the bipartite case, harvesting tripartite entanglement is possible with a single instantaneous switch of each detector.  
The absence of any bipartite negativity
implies the 3-detector system has harvested genuine GHZ-type tripartite entanglement following the interaction,
rather than W-type entanglement.
Moreover,  it is   possible to harvest  tripartite entanglement even if the first two detectors that switch cannot communicate,  provided both detectors are able to communicate with last detector.  This is quite unlike the case of two detectors (with three switchings), where communication between the detectors is required.

All of the above
indicates that it is easier to harvest tripartite entanglement than bipartite entanglement, commensurate with
 earlier work in $(1+1)$ dimensions
\cite{Lorek2014tripartite}, since there are looser constraints on the communication between the detectors.

Our results are also compatible with requirement that in order for two detectors to become entangled with delta switching, at least one must couple to the field twice \cite{Simidzija2018no-go}.  Since the $\pi$-tangle depends on the entanglement between one detector and the subsystem consisting of the other two (averaged over all three combinations), we can interpret the resulting tripartite entanglement by noting that the two-detector subsystem always interacts with the field twice.  From this perspective, it is not surprising that it can become entangled with the remaining detector, provided that it is not the last to interact.

Our approach for tripartite entanglement harvesting can be  extended to harvesting $N$-partite entanglement using $N$ detectors in a straightforward way, allowing one to probe  $N$-point correlations in a quantum field. It can also be
extended 
to many of the same problems that were explored in the bipartite case.
The effects of spacetime curvature and event horizons are of particular interest. The entanglement structure of a quantum field on a black hole spacetime \cite{henderson2018harvesting,Tjoa2020vaidya,Robbins:Amplification} 
induces several new effects, including entanglement shadows in the vicinity of the black hole, and entanglement amplification if the black hole rotates. It would be interesting to see the extent to which such effects are present in the multipartite case.\\

\noindent\emph{Acknowledgements}
This work was supported in part by the Natural Science and Engineering Research Council of Canada and by Asian Office of Aerospace Research and Development Grant No. FA2386-19-1-4077.

\bibliography{ref}

\clearpage
\appendix
\begin{widetext}

\begin{center}
{\large{\bf Supplementary Material}}    
\end{center}

\section{Reduced Density Matrix of the Detectors}

Here we show how the reduced density matrix   \eqref{eq:rhoABC} is computed.

Using  
\begin{align}
    \exp\lb[\hat \mu_{D}(T_D) \otimes \hat Y_{D}(T_D)\rb]
    &= \id_{D} \otimes \cosh \big(\hat Y_{D}(T_D)\big)
    +
    \hat \mu_{D}(T_D) \otimes \sinh \big(\hat Y_{D}(T_D)\big), 
\end{align}
 the nonperturbative expression   \eqref{eq:expanded unitary}  the time evolution operator   can be expanded as
\begin{align}
    \hat U\ts{I}
    &=
        \Bigkako{
            \id_{A}\otimes \id_{B} \otimes \id_{C}\otimes \cosh \hat Y_{C}
            +
            \id_{A}\otimes \id_{B} \otimes \hat m_{C}(T_C)\otimes \sinh \hat Y_{C}
        } \notag \\
        &\hspace{5mm}\times 
        \Bigkako{
            \id_{A}\otimes \id_{B} \otimes \id_{C}\otimes \cosh \hat Y_{B}
            +
            \id_{A}\otimes \hat m_{B}(T_B) \otimes \id_{C}\otimes \sinh \hat Y_{B}
        } \notag \\
        &\hspace{5mm}\times 
        \Bigkako{
            \id_{A}\otimes \id_{B} \otimes \id_{C}\otimes \cosh \hat Y_{A}
            +
            \hat m_{A}(T_A)\otimes \id_{B} \otimes \id_{C}\otimes \sinh \hat Y_{A}
        }, \label{eq:Utrig}
\end{align}
where we use the shorthand $\hat{Y}_D \coloneqq \hat{Y}_D(T_D)$.
This can be further simplified by using the identities $\cosh \hat Y_{D}=(e^{\hat Y_{D}} + e^{-\hat Y_{D}})/2$ and $\sinh \hat Y_{D}=(e^{\hat Y_{D}} - e^{-\hat Y_{D}})/2$ and noting that each term in Eq.\ \eqref{eq:Utrig} can be written as
\begin{align}
    \big(\hat{m}_A(T_{A})\big)^{(1-z)/2} \tp \big(\hat{m}_B(T_{B})\big)^{(1-y)/2} \big(\hat{m}_C(T_{C})\big)^{(1-x)/2} \tp \hat{X}_{(x,y,z)}
\end{align}
for $x,y,z\in \{ 1,-1 \}$ and where
\begin{align}
    \hat{X}_{(x,y,z)} &\coloneqq \frac{1}{2^3} \lb(\ec^{\hat{Y}_{C}} + x\ec^{-\hat{Y}_{C}}\rb) \lb(\ec^{\hat{Y}_{B}} + y\ec^{-\hat{Y}_{B}}\rb) \lb(\ec^{\hat{Y}_{A}} + z\ec^{-\hat{Y}_{A}}\rb). 
\end{align}
Therefore we have
\begin{align}
    \hat U\ts{I}
        &=
        \sum_{x,y,z} \big(\hat{m}_A(T_{A})\big)^{(1-z)/2} \tp \big(\hat{m}_B(T_{B})\big)^{(1-y)/2} \tp \big(\hat{m}_C(T_{C})\big)^{(1-x)/2} \tp \hat{X}_{(x,y,z)}.
	\label{eq:UX}
\end{align}

When the operator $\hat U\ts{I}$ is applied to the initial state of the detectors $\ket{0_{A} 0_{B} 0_{C}}$, the resulting state is
\begin{align}
    &\hat U\ts{I}\ket{0_{A} 0_{B} 0_{C}} \notag \\
    &= 
    \Bigg(\sum_{x,y,z} \big(\hat{m}_A(T_{A})\big)^{(1-z)/2} \big(\hat{m}_B(T_{B})\big)^{(1-y)/2} \tp \big(\hat{m}_C(T_{C})\big)^{(1-x)/2} \tp \hat{X}_{(x,y,z)}\Bigg) \Ket{0_A0_B0_C} \nn\\
	&= \sum_{x,y,z}\lb(\ec^{\iu\Omega_AT_{A}}\rb)^{(1-z)/2} \lb(\ec^{\iu\Omega_BT_{B}}\rb)^{(1-y)/2}  \lb(\ec^{\iu\Omega_CT_{C}}\rb)^{(1-x)/2} \Ket{\lb(\dfrac{1-z}{2}\rb)_A \lb(\dfrac{1-y}{2}\rb)_B \lb(\dfrac{1-x}{2}\rb)_C}  \tp \hat{X}_{(x,y,z)}.
\end{align}
From this, the reduced density matrix $\hat \rho_{ABC}$ is 
obtained by tracing over the field
\begin{align}
    \hat \rho_{ABC}
    &=
        \Tr_\phi [ \hat U\ts{I} \hat \rho_0 \hat U\ts{I}^\dag ] \notag \\
    &=
    \sum_{\substack{x,y,z\\ \qew,r,s}} \Bigg[\ec^{\iu(1-z)\Omega_AT_{A}/2}\ \ec^{\iu(1-y)\Omega_BT_{B}/2}\  \ec^{\iu(1-x)\Omega_CT_{C}/2}\ \ec^{-\iu(1-s)\Omega_AT_{A}/2}\ \ec^{-\iu(1-r)\Omega_BT_{B}/2}\ \ec^{-\iu(1-\qew)\Omega_CT_{C}/2} \nn\\
	&\hspace{5mm}\times  \Braket{0|\hat{X}_{(\qew,r,s)}^\dagger \hat{X}_{(x,y,z)}|0}\;   \Ket{\lb(\dfrac{1-z}{2}\rb)_A \lb(\dfrac{1-y}{2}\rb)_B \lb(\dfrac{1-x}{2}\rb)_C}\Bra{\lb(\dfrac{1-s}{2}\rb)_A \lb(\dfrac{1-r}{2}\rb)_B \lb(\dfrac{1-\qew}{2}\rb)_C} \Bigg],
\end{align}
using  the fact that the smeared field is anti-Hermitian, $\hat Y_{D}^\dag = -\hat Y_{D}$.

To obtain the final expression \eqref{eq:rhoABC}, let us focus on $\bra{0} \hat{X}_{(\qew,r,s)}^\dagger \hat{X}_{(x,y,z)} \ket{0}$.
Using the result
\cite{Simidzija.Nonperturbative}
\begin{align}
    \Braket{0|\ec^{u\hat{Y}_{D}+v\hat{Y}_{E}}|0} = \exp\lb(-\frac{1}{2} \int d^n\bd{k}\ \big|u\beta_{D}(\bd{k})+v\beta_{E}(\bd{k})\big|^2\rb),
\end{align}
for $u,v\in\mathbb{C}$ and $\beta_D(\bd{k})$ defined in \eqref{eq:beta}, we employ the Baker-Campbell-Hausdorff formula to obtain
\begin{align}
    &
    \Braket{0|\hat{X}_{(\qew,r,s)}^\dagger \hat{X}_{(x,y,z)}|0} \nn\\
	&\quad = 
	\dfrac{1}{2^6} \bigg<0\ \bigg| \lb(\ec^{-\hat{Y}_{A}} + s\ec^{\hat{Y}_{A}}\rb) \lb(\ec^{-\hat{Y}_{B}} + r\ec^{\hat{Y}_{B}}\rb) \lb(\ec^{-\hat{Y}_{C}} + \qew\ec^{\hat{Y}_{C}}\rb) \lb(\ec^{\hat{Y}_{C}} + x\ec^{-\hat{Y}_{C}}\rb) \lb(\ec^{\hat{Y}_{B}} + y\ec^{-\hat{Y}_{B}}\rb) \lb(\ec^{\hat{Y}_{A}} + z\ec^{-\hat{Y}_{A}}\rb)\bigg|\ 0\bigg> \nn\\
	&\quad = 
	\dfrac{1}{64} \sum_{\substack{a,b,c\\j,k,\ell}} s^{(1+a)/2} r^{(1+b)/2} \qew^{(1+c)/2} x^{(1-j)/2} y^{(1-k)/2} z^{(1-\ell)/2} \Braket{0| \ec^{a\hat{Y}_{A}} \ec^{b\hat{Y}_{B}} \ec^{c\hat{Y}_{C}} \ec^{j\hat{Y}_{C}} \ec^{k\hat{Y}_{B}} \ec^{\ell\hat{Y}_{A}} |0} \nn\\
	&\quad = 
	\dfrac{1}{64} \sum_{\substack{a,b,c\\j,k,\ell}} s^{(1+a)/2} r^{(1+b)/2} \qew^{(1+c)/2} x^{(1-j)/2} y^{(1-k)/2} z^{(1-\ell)/2}\  \Big(f_{A}^{(a+\ell)^2} f_{B}^{(b+k)^2} f_{C}^{(c+j)^2} \ec^{(a+\ell)(b+k)\omega_{A,B}}\ \ec^{(a+\ell)(c+j)\omega_{A,C}} \nn\\
	&\hspace{10mm} \times \ec^{(b+k)(c+j)\omega_{B,C}}\ \ec^{\iu/2(a-\ell)(b+k)\Theta_{A,B}}\ \ec^{\iu/2(a-\ell)(c+j)\Theta_{A,C}}\ \ec^{\iu/2(b-k)(c+j)\Theta_{B,C}}\Big)
\end{align}
for $a,b,c,j,k,\ell \in \Set{-1,+1}$
and where $f_D$, $\Theta_{D,E}$ and $\omega_{D,E}$ are defined in equation \ref{eq:MatrixFunctions}.

Finally, we arrive at the expression for the density matrix describing the final state of the three detectors:
\begin{align}
	\hat{\rho}_{ABC} 
	&= 
	\dfrac{1}{64} \sum_{\substack{x,y,z\\ \qew,r,s}} \Bigg[\ec^{\iu(1-z)\Omega_AT_{A}/2}\ \ec^{\iu(1-y)\Omega_BT_{B}/2}\  \ec^{\iu(1-x)\Omega_CT_{C}/2}\ \ec^{-\iu(1-s)\Omega_AT_{A}/2}\ \ec^{-\iu(1-r)\Omega_BT_{B}/2}\ \ec^{-\iu(1-\qew)\Omega_CT_{C}/2} \nn\\
	&\hspace{5mm} \times \sum_{\substack{a,b,c\\j,k,\ell}} \Big(s^{(1+a)/2} r^{(1+b)/2} \qew^{(1+c)/2} x^{(1-j)/2} y^{(1-k)/2} z^{(1-\ell)/2} f_{A}^{(a+\ell)^2} f_{B}^{(b+k)^2} f_{C}^{(c+j)^2}\ \ec^{(a+\ell)(b+k)\omega_{A,B}}
\nn\\
	&\hspace{15mm} \times \ec^{(a+\ell)(c+j)\omega_{A,C}}\ \ec^{(b+k)(c+j)\omega_{C}}\ \ec^{\iu/2(a-\ell)(b+k)\Theta_{A,B}}\ \ec^{\iu/2(a-\ell)(c+j)\Theta_{A,C}}\ \ec^{\iu/2(b-k)(c+j)\Theta_{B,C}}\Big) \nn\\
	&\hspace{5mm} \times \Ket{\lb(\frac{1-z}{2}\rb)_A \lb(\frac{1-y}{2}\rb)_B \lb(\frac{1-x}{2}\rb)_C}\Bra{\lb(\frac{1-s}{2}\rb)_A \lb(\frac{1-r}{2}\rb)_B \lb(\frac{1-\qew}{2}\rb)_C} \Bigg]
	\label{eq:rhoABCother}
\end{align}
where
	$a,b,c,j,k,\ell,\qew,r,s,x,y,z \in \Set{-1,+1}$.
Evaluating the sums results the density matrix   \eqref{eq:rhoABC}, where the individual elements can be read off in the basis $\{ \ket{0_{A}0_{B}0_{C}}, \ket{0_{A}0_{B}1_{C}}$, $\ket{0_{A}1_{B}0_{C}}, \ket{1_{A}0_{B}0_{C}}$, $\ket{0_{A}1_{B}1_{C}}$, $\ket{1_{A}0_{B}1_{C}}$, $\ket{1_{A}1_{B}0_{C}}, \ket{1_{A}1_{B}1_{C}} \}$.
 For completeness,  we have included some of the elements of the density matrix   \eqref{eq:rhoABC}:
\begin{subequations}
    \begin{align}
        &r_{11}
    =
        \dfrac{1}{8}
        \biggkako{
            1 
            + f_{A}^4
            + f_{B}^4 \cos (2\Theta_{AB})
            + f_{A}^4 f_{B}^4 \cosh (4\omega_{AB}) \notag \\
            &
            + f_{C}^4 \cos (2\Theta_{AC}) \cos (2\Theta_{BC}) 
            + f_{A}^4 f_{C}^4 \cos(2\Theta_{BC}) \cosh(4\omega_{AC}) \notag \\
            &
            + f_{B}^4 f_{C}^4
            \Bigkagikako{
                \cos(2\Theta_{AB}) \cos (2\Theta_{AC}) \cosh(4\omega_{BC})
                -
                \sin(2\Theta_{AB}) \sin (2\Theta_{AC}) \sinh(4\omega_{BC})
            } \notag \\
            &
            +
            f_{A}^4 f_{B}^4 f_{C}^4
            \Bigkagikako{
                \cosh (4\omega_{AC}) \cosh (4\omega_{BC}) \cosh (4\omega_{AB})
                +
                \sinh (4\omega_{AC}) \sinh (4\omega_{BC}) \sinh (4\omega_{AB})
            }
        }, \\
    &r_{22}
    =
        \dfrac{1}{8}
        \biggkako{
            1 
            + f_{A}^4
            + f_{B}^4 \cos (2\Theta_{AB})
            + f_{A}^4 f_{B}^4 \cosh (4\omega_{AB}) \notag \\
            &
            - f_{C}^4 \cos (2\Theta_{AC}) \cos (2\Theta_{BC}) 
            - f_{A}^4 f_{C}^4 \cos(2\Theta_{BC}) \cosh(4\omega_{AC}) \notag \\
            &
            - f_{B}^4 f_{C}^4
            \Bigkagikako{
                \cos(2\Theta_{AB}) \cos (2\Theta_{AC}) \cosh(4\omega_{BC})
                -
                \sin(2\Theta_{AB}) \sin (2\Theta_{AC}) \sinh(4\omega_{BC})
            } \notag \\
            &
            -
            f_{A}^4 f_{B}^4 f_{C}^4
            \Bigkagikako{
                \cosh (4\omega_{AC}) \cosh (4\omega_{BC}) \cosh (4\omega_{AB})
                +
                \sinh (4\omega_{AC}) \sinh (4\omega_{BC}) \sinh (4\omega_{AB})
            }
        }, \\
        &r_{15}
    =
        \dfrac{1}{8}
        \ec^{ -\ii (\Omega_{B}T_{B} + \Omega_{C} T_{C}) }
        f_{C}^4
        \biggkako{
            \ii \cos(2\Theta_{AC}) \sin (2\Theta_{BC})
            +
            \ii f_{A}^4 \cosh(4\omega_{AC}) \sin(2\Theta_{BC}) \notag \\
            &+
            f_{B}^4 
            \Bigkagikako{
                \cos(2\Theta_{AB}) \cos(2\Theta_{AC}) \sinh(4\omega_{BC})
                -
                \sin(2\Theta_{AB}) \sin(2\Theta_{AC}) \cosh(4\omega_{BC})
            } \notag \\
            &+
            f_{A}^4 f_{B}^4
            \Bigkagikako{
                \cosh(4\omega_{BC}) \sinh(4\omega_{AB}) \sinh(4\omega_{AC})
                +
                \sinh(4\omega_{BC}) \cosh(4\omega_{AB}) \cosh(4\omega_{AC})
            }
        },
    \end{align}
\end{subequations}
and the remaining elements can be similarly found from Eq.\ \eqref{eq:rhoABCother}.

\end{widetext}

\end{document}